\documentclass[a4paper,11pt]{article}
\usepackage{jheppub} 
\usepackage[T1]{fontenc}

\usepackage{amsmath,amssymb,graphicx,hyperref}

\usepackage{tikz} 
\usetikzlibrary{shapes,arrows,positioning,automata,backgrounds,calc,er,patterns,  shapes.geometric}
\usetikzlibrary{calc,arrows.meta,positioning}
\usepackage{tikz-feynman}
\tikzfeynmanset{compat=1.1.0}
\usepackage{graphicx}

\title{Artificial Intelligence and Symmetries: \\ 
Learning, Encoding, and Discovering Structure in Physical Data}

\author{Veronica Sanz\footnote{email: veronica.sanz@uv.es}}
\affiliation{Instituto de F\'isica Corpuscular (IFIC), Universitat de Val\`encia--CSIC,\\
E-46980 Val\`encia, Spain}

\abstract{
Symmetries play a central role in physics, organizing dynamics, constraining interactions,
and determining the effective number of physical degrees of freedom. In parallel, modern
artificial intelligence methods have demonstrated a remarkable ability to extract
low-dimensional structure from high-dimensional data through representation learning.
This review examines the interplay between these two perspectives, focusing on the extent
to which symmetry-induced constraints can be identified, encoded, or diagnosed using
machine learning techniques.

Rather than emphasizing architectures that enforce known symmetries by construction,
we concentrate on data-driven approaches and latent representation learning, with
particular attention to variational autoencoders. We discuss how symmetries and
conservation laws reduce the intrinsic dimensionality of physical datasets, and how this
reduction may manifest itself through self-organization of latent spaces in generative
models trained to balance reconstruction and compression. We review recent results,
including case studies from simple geometric systems and particle physics processes, and
analyze the theoretical and practical limitations of inferring symmetry structure without
explicit inductive bias.

{\it This manuscript is an invited review at the International Journal of Modern Physics A.}
}

\begin{document}
\maketitle
\flushbottom

\section{Introduction}
\label{sec:introduction}

Symmetries play a central role in modern theoretical physics. They organize physical
laws, constrain dynamics, and determine which degrees of freedom are physically
relevant. From Noether’s theorem, which links continuous symmetries to conservation
laws, to gauge invariance and spacetime symmetries in quantum field theory, much of
our understanding of fundamental interactions can be traced back to symmetry
principles \cite{Noether1918,WeinbergQFT1,WeinbergQFT2}. In this sense, symmetries
are not merely aesthetic features of physical theories, but constitute a form of
information compression: they encode redundancy in the description of physical
systems and identify equivalence classes of configurations that are physically
indistinguishable.

Traditionally, symmetry considerations enter physics through analytic reasoning and
model building. One postulates a symmetry group, constrains the allowed operators or
dynamics accordingly, and confronts the resulting theory with data. This approach has
been extraordinarily successful, yet it relies on a crucial assumption: that the
relevant symmetries are known \emph{a priori}. In complex systems, high-dimensional
datasets, or situations where symmetries are only approximate, emergent, or partially
broken, this assumption may no longer be justified. This raises a natural question:
\emph{can symmetries be inferred directly from data, without being imposed by hand?}

Recent advances in artificial intelligence, and in particular in representation learning,
have brought renewed attention to this question. Modern machine learning models are
capable of extracting low-dimensional structure from extremely high-dimensional data,
often discovering latent representations that are more compact, structured, and
interpretable than the original input space \cite{BengioRepLearning,GoodfellowDL}.
In scientific contexts, this ability has been exploited for pattern recognition,
acceleration of simulations, and surrogate modeling. More subtly, it has also opened
the possibility that learning algorithms may act as \emph{diagnostic tools}, revealing
hidden constraints, redundancies, or organizing principles in data.

The relationship between machine learning and symmetry has so far developed along
three partially distinct directions. First, known symmetries can be explicitly enforced
at the architectural level. Convolutional neural networks exploit translational
invariance through weight sharing \cite{LeCunCNN}, while more general group-equivariant
and geometric deep learning frameworks encode rotational, permutation, Lorentz, or
gauge symmetries directly into network layers
\cite{CohenWelling2016,BronsteinGDL}. When the symmetry group and its action on the
data are known, this approach yields clear benefits in sample efficiency,
generalization, and interpretability.

Second, symmetries can be encouraged indirectly through data augmentation or
self-supervised objectives. By exposing a model to symmetry-related versions of the
same data point—rotations, boosts, or other transformations—one can bias the learned
representation toward invariance or equivariance, even if this is not enforced exactly
by the architecture \cite{ChenSimCLR,GrillBYOL}. While powerful, this strategy still
presupposes some knowledge of the relevant transformations.

A third, more exploratory direction asks whether symmetry-related structure can
\emph{emerge spontaneously} in the internal representations of learning algorithms,
even when no symmetry is imposed or labeled. Early evidence for this phenomenon
appeared in studies of disentangled representation learning, where unsupervised
generative models were shown to align latent variables with human-interpretable
factors of variation such as position, orientation, or scale
\cite{HigginsBetaVAE,InfoGAN}. Subsequent work, however, clarified that purely
unsupervised disentanglement is not identifiable in general, and cannot be guaranteed
without inductive biases or supervision \cite{Locatello2019}. This result underscores
the need for caution: emergent structure in latent space should be treated as a
diagnostic, not as a proof of underlying physical factors.

From the perspective of physics, this diagnostic viewpoint is nonetheless extremely
appealing. Physical symmetries imply constraints and redundancies that reduce the
effective dimensionality of observable data. A learning algorithm trained to compress
and reconstruct data efficiently should therefore be incentivized to allocate
information preferentially along symmetry-independent directions, while suppressing
redundant ones. If this intuition is correct, the organization of latent representations
may provide a data-driven window into the symmetry structure of the underlying system,
even when that structure is only approximate or partially obscured.

The purpose of this review is to critically examine this idea. Rather than focusing on
architectures that enforce symmetry by construction, we concentrate on standard
representation-learning models and ask under which conditions symmetry-induced
structure manifests itself in their latent spaces. Particular emphasis is placed on
variational autoencoders (VAEs), which combine nonlinear dimensional reduction with a
well-defined probabilistic notion of compression \cite{KingmaWelling}. We review how
symmetries, constraints, and conservation laws can lead to latent-space
self-organization, and how this organization can be quantified and interpreted.

This article is based on a sequence of recent works~\cite{Sanz:2025sld,Barenboim:2021vzh,barenboim2024exploring} 
 and invited talks, but its scope
is deliberately broader. Our goal is not to advocate a specific algorithm as a universal
symmetry detector, but to clarify what can—and cannot—be inferred from data-driven
latent representations. Throughout, we emphasize limitations as much as successes,
and we frame machine learning as a complementary tool to traditional theoretical
reasoning, rather than as a replacement for it.

\begin{figure}[t]
\centering
\begin{tikzpicture}[scale=0.95]

\draw[thick] (-4,0) rectangle (-1,3);
\node at (-2.5,3.3) {\small Data space};

\foreach \y in {0.6,1.5,2.4} {
  \draw[gray!60] (-2.5,\y) ellipse (0.9 and 0.25);
  \fill[gray!60] (-3.2,\y) circle (1.2pt);
  \fill[gray!60] (-2.0,\y) circle (1.2pt);
}

\node[align=center, gray!70] at (-2.5,-0.6)
{\footnotesize Redundant\\coordinates};

\draw[->, thick] (-0.8,1.5) -- (0.2,1.5);

\draw[thick] (0.4,0.4) to[out=90,in=180] (1.8,2.6)
             to[out=0,in=90] (3.2,0.4);

\node at (1.8,3.3) {\small Constraint manifold};

\draw[->, gray!70] (1.3,1.6) -- (1.6,1.9);
\draw[->, gray!70] (2.3,1.2) -- (2.6,1.5);

\node[align=center, gray!70] at (1.8,-0.6)
{\footnotesize Symmetry\\orbits};

\draw[->, thick] (3.6,1.5) -- (4.6,1.5);

\draw[->, thick] (4.9,0.6) -- (6.4,0.6);
\draw[->, thick] (4.9,0.6) -- (4.9,2.1);
\draw[->, thick] (4.9,0.6) -- (5.8,1.9);

\node at (6.6,0.5) {\small $z_1$};
\node at (4.7,2.3) {\small $z_2$};
\node at (5.9,2.2) {\small $z_3$};

\draw[very thick, blue] (4.9,0.6) -- (6.4,0.6);

\draw[thick, gray!40] (4.9,0.6) -- (4.9,2.1);
\draw[thick, gray!40] (4.9,0.6) -- (5.8,1.9);

\node[align=center] at (5.7,3.3) {\small Latent space};

\node[align=center, gray!70] at (5.7,-0.6)
{\footnotesize Effective\\degrees of freedom};

\end{tikzpicture}

\caption{Schematic view of the role of symmetries in data-driven representation learning.
Physical symmetries induce constraints and redundancies in the data manifold, reducing
the effective number of independent degrees of freedom. Representation-learning models
trained to balance reconstruction and compression may reflect this structure through
self-organization of their latent spaces.}
\label{fig:symmetry_ai_schematic}
\end{figure}
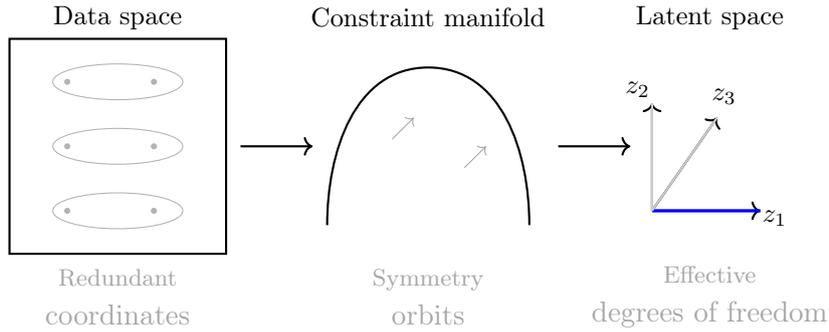

\section{Symmetries as Information and Constraints}
\label{sec:symmetry_information}

In physics, symmetries are not merely transformations under which equations remain
invariant; they encode fundamental information about what distinguishes physically
inequivalent states from redundant descriptions. Two configurations related by a symmetry
transformation correspond to the same physical situation, even though they may differ at
the level of coordinates or fields. In this sense, symmetries identify equivalence classes
in the space of descriptions and thereby reduce the effective number of independent
degrees of freedom.

This observation can be formalized in a variety of familiar contexts. In classical
mechanics, invariance under spatial translations implies conservation of momentum, while
rotational invariance implies conservation of angular momentum, as formalized by
Noether’s theorem \cite{Noether1918}. In quantum field theory, internal symmetries constrain
the allowed interactions and operator content of a theory, while spacetime symmetries
govern dispersion relations, selection rules, and the structure of scattering amplitudes
\cite{WeinbergQFT1,WeinbergQFT2}. In all these cases, symmetry reduces arbitrariness: it
forbids certain variations while identifying others as physically irrelevant.

From an information-theoretic perspective, this role of symmetry can be viewed as a form
of redundancy removal. If a dataset is generated by a physical process with an exact
symmetry, then multiple data points related by that symmetry carry no additional
independent information. The true information content of the dataset resides in variables
that parametrize inequivalent symmetry orbits, rather than in the raw coordinates
themselves. The dimension of this reduced space is often referred to as the
\emph{intrinsic dimensionality} of the system.

A simple geometric example illustrates this point. Consider data points distributed on a
circle embedded in a two-dimensional space. Although each data point is specified by two
coordinates, the physical configuration is fully described by a single angular variable.
The continuous rotational symmetry identifies all points related by a global rotation as
equivalent up to that angle. The constraint defining the circle, together with the
symmetry, reduces the effective degrees of freedom from two to one. Similar reasoning
applies to more complex manifolds, group orbits, and constrained dynamical systems.

In many physical systems, symmetries manifest themselves not only through invariance but
also through explicit constraints. Conservation laws impose relations among observables
that must hold event by event, reducing the dimensionality of the space of allowed
configurations. In particle collisions, for instance, energy--momentum conservation
relates the momenta of final-state particles, while on-shell conditions impose additional
constraints \cite{PeskinSchroeder}. These relations define a lower-dimensional manifold
within the space of all kinematically allowed variables.

Importantly, not all symmetries encountered in practice are exact. Approximate symmetries,
emergent symmetries, and explicitly broken symmetries are ubiquitous in physics.
Chiral symmetry in QCD, Lorentz symmetry in effective theories with cutoffs, or approximate
conservation laws arising from scale separation all provide examples where symmetry holds
only within a limited regime of validity \cite{GeorgiEFT}. In such cases, symmetry-induced
constraints still shape the structure of the data, but in a softened or distorted manner.
The effective dimensionality may be reduced only approximately, and deviations from exact
constraints carry physically meaningful information.

This distinction is crucial when considering data-driven analyses. A learning algorithm
trained on data generated by a symmetric system is not exposed to the abstract symmetry
group itself, but only to its concrete consequences: correlations, constraints, and
redundancies among observables. From this viewpoint, detecting symmetry amounts to
detecting a structured reduction of information. Exact symmetries correspond to sharp
constraints and clear dimensional reduction, while approximate symmetries manifest as
hierarchies, soft modes, or preferred directions in the space of variations.

\begin{figure}[ht!]
\centering
\includegraphics[width=0.9 \textwidth]{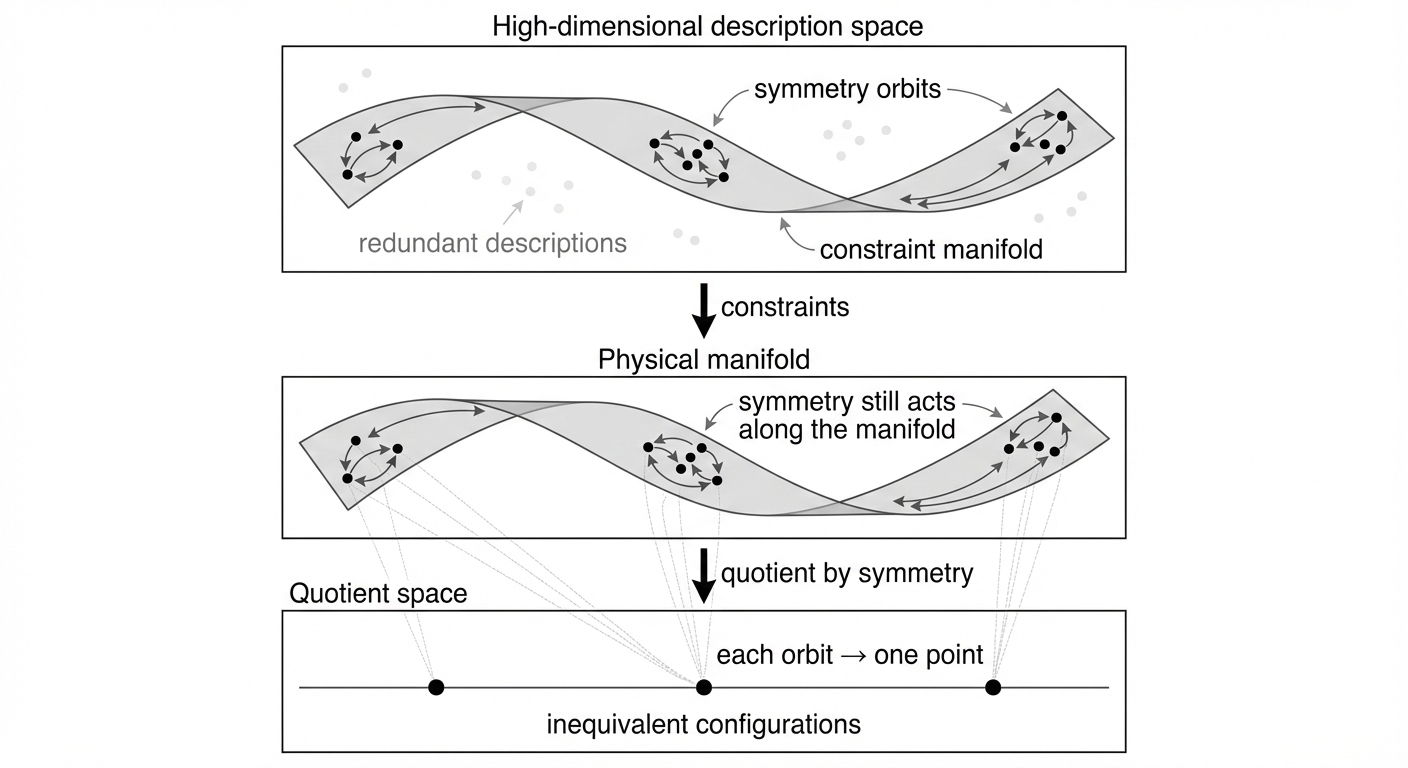}
\caption{Illustration of symmetry as information reduction. A high-dimensional space of
descriptions (top) contains redundancy due to symmetry transformations, which identify
equivalence classes of physically identical configurations. Constraints and conservation
laws restrict the data to a lower-dimensional manifold, while symmetry further quotients
this manifold into physically inequivalent degrees of freedom (bottom). {\it Figure generated with AI.}}
\label{fig:symmetry_information}
\end{figure}

The central theme of this review is that these information-theoretic signatures of symmetry
can, under suitable conditions, be reflected in learned representations of data.
Representation-learning models trained to balance fidelity and compression are naturally
sensitive to redundancy. If a dataset contains symmetry-related degeneracies, an efficient
representation should suppress redundant directions and allocate capacity to variables
that parametrize physically distinct configurations. The extent to which this happens,
and how reliably it can be diagnosed, is the subject of the sections that follow.


\section{Symmetry in Machine Learning: Three Paradigms}
\label{sec:symmetry_ml}

The interaction between symmetry and machine learning has developed along several
conceptually distinct lines. While these approaches are often grouped together under the
broad label of ``symmetry-aware learning'', they differ substantially in their assumptions,
objectives, and epistemic status. In this section, we organize the existing literature into
three paradigms, ordered by the degree to which symmetry is specified \emph{a priori}.
This classification will be useful for clarifying what can reasonably be expected from
data-driven symmetry diagnostics in later sections.

\subsection{Architectural Symmetry: Invariance and Equivariance}
\label{subsec:architectural_symmetry}

The most direct way to incorporate symmetry into a learning model is to encode it
explicitly at the architectural level. This approach mirrors the traditional theoretical
strategy in physics: one identifies the relevant symmetry group and restricts the space of
allowed functions accordingly. In machine learning, this idea appears as invariance or
equivariance under group actions.

The canonical example is the convolutional neural network (CNN), where translational
invariance is implemented through weight sharing and local connectivity
\cite{LeCunCNN}. This principle has since been generalized to arbitrary groups, leading
to group-equivariant convolutional networks and, more broadly, to the field of geometric
deep learning \cite{CohenWelling2016,BronsteinGDL}. In particle physics, these ideas have
been extended to permutation-invariant architectures for sets of particles, Lorentz-
equivariant networks, and models respecting gauge or spacetime symmetries
\cite{BogatskiyPELICAN,HaoLorentzAE}.

When the symmetry group and its action on the data are known, architectural symmetry
provides strong guarantees. The hypothesis space is restricted to symmetry-consistent
functions, reducing sample complexity and improving generalization. From a physics
perspective, this approach is conceptually clean: the symmetry principle is imposed
before learning, much like in effective field theory or model building.

However, this strength is also its main limitation. Architectural symmetry requires
\emph{prior knowledge} of the relevant group, its representation, and how it acts on the
data. It is therefore ill-suited to situations where symmetries are unknown, approximate,
emergent, or explicitly broken, or where the correct variables on which the symmetry acts
are themselves unclear. In such cases, enforcing an incorrect symmetry can bias the model
and obscure physically relevant structure.

\subsection{Implicit Symmetry via Data Augmentation and Self-Supervision}
\label{subsec:implicit_symmetry}

A second paradigm incorporates symmetry indirectly, without enforcing it exactly at the
level of the architecture. Instead, symmetry is introduced through the training procedure,
most commonly via data augmentation or self-supervised objectives. The model is exposed
to multiple transformed versions of the same data point and is encouraged to produce
similar representations for all of them.

This strategy has been particularly influential in representation learning for images and
signals, where contrastive and self-supervised methods use augmentations such as
translations, rotations, or color transformations to learn invariant features
\cite{ChenSimCLR,GrillBYOL}. In physics-inspired applications, analogous ideas appear in
the use of Lorentz boosts, rotations, or permutations as augmentation strategies
\cite{JetCLR,DillonSymmetrySafety}.

Implicit symmetry has two notable advantages. First, it is flexible: approximate or
domain-specific symmetries can be incorporated without requiring an exact group-theoretic
formulation. Second, it allows the same architecture to be reused across different
symmetry assumptions. However, this flexibility comes at the cost of reduced guarantees.
The learned invariance is only as good as the augmentation scheme, and different choices
of transformations can lead to qualitatively different representations.

From a conceptual standpoint, this paradigm still relies on prior knowledge of symmetry.
The transformations used for augmentation are chosen by the practitioner, and the model
is guided toward invariance by construction. As such, implicit symmetry learning does not
constitute symmetry discovery, but rather symmetry \emph{enforcement through training}.

\subsection{Emergent Structure in Latent Representations}
\label{subsec:emergent_symmetry}

The third paradigm is more exploratory and is the primary focus of this review. Here,
symmetry is neither imposed architecturally nor encoded explicitly through data
augmentation. Instead, one asks whether symmetry-related structure can emerge
spontaneously in the internal representations of learning algorithms trained on raw data.

This question arose prominently in the context of unsupervised and weakly supervised
representation learning. Early work on disentangled representations suggested that latent
variables in generative models could align with interpretable factors of variation, such as
position, orientation, or scale, without explicit supervision \cite{HigginsBetaVAE,InfoGAN}. Subsequent analyses, however, demonstrated that such alignment is not
identifiable in general and cannot be guaranteed without inductive biases or supervision
\cite{Locatello2019}. These results impose a fundamental limitation: emergent structure in
latent space must be interpreted diagnostically, not ontologically.
\begin{figure}[t]
\centering
\includegraphics[width=0.95\textwidth]{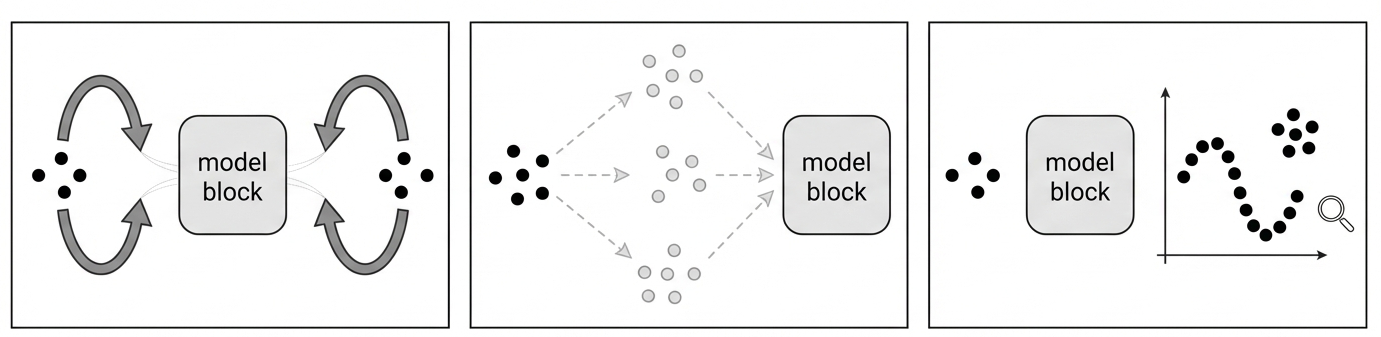}
\caption{Three paradigms for incorporating symmetry in machine learning.
\emph{Left}: Architectural symmetry, where invariance or equivariance is imposed by
construction.
\emph{Center}: Implicit symmetry, introduced through data augmentation or self-supervised
objectives.
\emph{Right}: Emergent structure, where symmetry-related organization may arise in latent
representations without explicit enforcement, and is interpreted diagnostically. {\it Figure generated with AI.}}
\label{fig:symmetry_paradigms}
\end{figure}

Despite this limitation, emergent latent organization remains of significant interest for
physics. Physical symmetries do not merely correspond to abstract transformations; they
impose concrete constraints and redundancies in observable data. A learning algorithm
trained to compress data efficiently is therefore exposed to the \emph{consequences} of
symmetry, even if it is ignorant of the symmetry itself. If the data contain exact or
approximate degeneracies, an efficient representation may suppress redundant directions
and allocate representational capacity to symmetry-independent degrees of freedom.

In this paradigm, the goal is not to recover a symmetry group or its generators explicitly,
but to detect structured reductions of dimensionality, hierarchies of relevance, or
preferred directions in representation space. These features can serve as indicators of
underlying constraints or approximate symmetries, provided their limitations are clearly
understood. The remainder of this review is devoted to examining this diagnostic approach
in detail, with particular emphasis on variational autoencoders as a concrete and
interpretable framework.

\section{Variational Autoencoders as Probes of Symmetry}
\label{sec:vae_symmetry}

The previous sections emphasized that physical symmetries manifest themselves as
redundancies, constraints, and reductions in effective dimensionality of observable data.
If symmetry is to be diagnosed directly from data, rather than imposed by construction,
the learning algorithm must satisfy two basic requirements. First, it must learn a
compressed representation that reflects the intrinsic structure of the data, rather than
simply memorizing it. Second, it must provide a principled way of quantifying which
directions in representation space are meaningfully used and which are redundant. In this
section we argue that variational autoencoders (VAEs) provide a natural framework for
addressing these requirements.

\subsection{Compression, Reconstruction, and Latent Variables}

A variational autoencoder is a generative model that learns a probabilistic mapping between
observed data $x$ and a set of latent variables $z$, trained by maximizing the evidence
lower bound (ELBO) \cite{KingmaWelling}. The objective balances two competing goals:
accurate reconstruction of the data and compression of the latent representation toward a
simple prior distribution. Schematically,
\begin{equation}
\mathcal{L}_{\mathrm{VAE}}
=
\mathbb{E}_{q(z|x)}[\log p(x|z)]
-
\beta\, D_{\mathrm{KL}}(q(z|x)\,\|\,p(z)) ,
\end{equation}
where $q(z|x)$ is the encoder, $p(x|z)$ the decoder, and $p(z)$ a chosen prior, typically a
factorized Gaussian.

From the perspective of symmetry diagnostics, the crucial feature of this objective is not
its generative nature per se, but the explicit tension between reconstruction fidelity and
information compression. Directions in latent space that carry little information about
the data are penalized by the Kullback–Leibler term and driven toward the prior. Conversely,
latent directions that encode genuine variation across the dataset are retained. This
mechanism naturally mirrors the effect of symmetry-induced redundancy: if multiple data
points are related by constraints or symmetries, efficiently encoding all of them requires
fewer independent latent variables.

\subsection{Why VAEs Are Different from Other Dimensionality Reduction Methods}

At first sight, this logic resembles classical dimensionality reduction techniques such as
principal component analysis (PCA). Indeed, in linear settings, autoencoders and PCA are
closely related \cite{BaldiHornik}. However, several features distinguish VAEs as probes of
symmetry-induced structure.

First, VAEs operate in a nonlinear regime. Physical constraints and symmetry orbits often
define curved, non-Euclidean manifolds embedded in high-dimensional spaces. Linear methods
may detect reduced rank locally, but they generally fail to capture global structure. VAEs
can, in principle, learn nonlinear coordinate systems adapted to such manifolds.

Second, VAEs provide an explicit probabilistic latent representation. Each latent variable
is characterized not only by a mean value, but also by an event-wise uncertainty. This
distinction between variation across the dataset and uncertainty within a single data
point plays a central role in assessing which latent directions are genuinely informative.
Deterministic autoencoders lack a canonical notion of latent scale or relevance, making
post-hoc interpretation ambiguous.

Third, the prior imposed on the latent variables provides a reference against which
information content can be measured. Latent directions that are unused collapse to the
prior in a controlled way, rather than remaining arbitrarily scaled or entangled. This
property is essential for defining meaningful diagnostics of effective dimensionality and
latent-space organization.

\subsection{Relation to Disentanglement and Its Limitations}

Variational autoencoders have often been discussed in the context of disentangled
representation learning. Modifications of the standard objective, such as the $\beta$-VAE,
FactorVAE, or $\beta$-TCVAE, explicitly encourage statistical independence among latent
variables \cite{HigginsBetaVAE,KimMnih,ChenTCVAE}. While these approaches can promote
alignment between latent variables and independent factors of variation, it is now well
established that purely unsupervised disentanglement is not identifiable without inductive
bias or supervision \cite{Locatello2019}.

For the purposes of symmetry diagnostics, this result has an important implication.
Alignment of latent variables with symmetry directions should not be interpreted as the
recovery of a unique or physically privileged coordinate system. Instead, it should be
understood as evidence that the data admit a reduced and structured representation,
consistent with symmetry-induced redundancy. The goal is therefore not disentanglement in
the strong sense, but the identification of robust hierarchies or reductions in latent
space usage.

\subsection{Symmetry as Latent-Space Self-Organization}

When a VAE is trained on data generated by a physical system with exact or approximate
symmetries, it is exposed only to the observable consequences of those symmetries: reduced
degrees of freedom, correlations among variables, and constraints that hold across events.
If the model successfully balances reconstruction and compression, these features can
manifest themselves as a self-organization of the latent space.
\begin{figure}[t]
\centering
\includegraphics[width=0.85\textwidth]{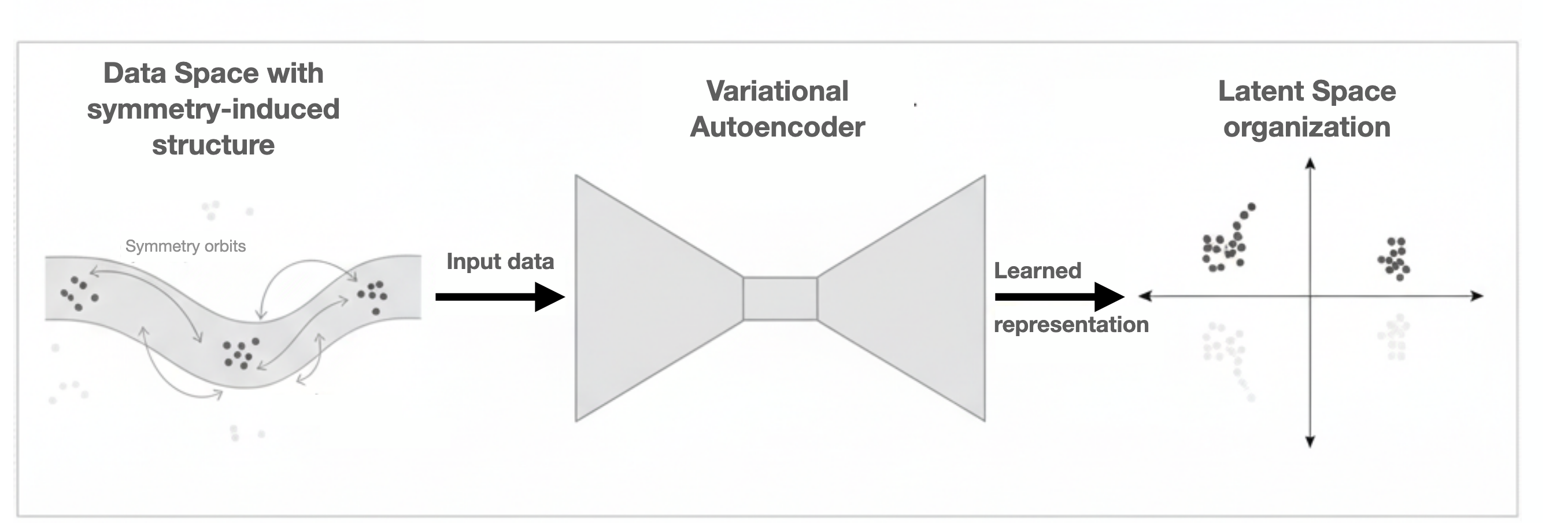}
\caption{Conceptual view of a variational autoencoder as a symmetry probe. Symmetry-induced
constraints restrict the data to a lower-dimensional manifold. During training, the VAE
balances reconstruction accuracy against compression toward a simple latent prior,
leading to suppression of redundant latent directions and preferential use of symmetry-
independent degrees of freedom.}
\label{fig:vae_symmetry_probe}
\end{figure}

In particular, one may observe that only a subset of latent variables carries significant
variation across the dataset, while the remaining directions are suppressed toward the
prior. The number of such relevant latent directions provides a data-driven estimate of the
effective dimensionality of the system. Moreover, correlations between latent variables and
specific combinations of input features can reveal how the model internally encodes
constraints such as conservation laws.

Crucially, this behavior is not guaranteed. It depends on the strength and clarity of the
symmetry-induced redundancy, the choice of architecture and hyperparameters, and the degree
to which the symmetry is exact or approximate. For this reason, VAEs should be viewed as
\emph{probes} of symmetry rather than symmetry detectors. They provide a way to test whether
the data themselves support a reduced, structured representation, without imposing that
structure by hand.

The remainder of this review is devoted to making this diagnostic viewpoint concrete. In
the next section, we introduce quantitative measures of latent relevance and illustrate
how latent-space self-organization emerges in simple toy models and in realistic examples
from particle physics.

\section{Latent-Space Self-Organization and Relevance Measures}
\label{sec:latent_relevance}

A recurring theme in the preceding sections is that physical symmetries manifest
themselves as redundancies and constraints in observable data. When a representation-
learning model is trained to compress such data efficiently, these redundancies may
translate into a nontrivial organization of the latent space. In this section, we make
this statement precise by introducing quantitative diagnostics that allow one to assess
how many latent directions are effectively used, and how this usage reflects symmetry-
induced structure.

\subsection{Self-Organization in Overparameterized Latent Spaces}

Throughout this review, we focus on deliberately \emph{overparameterized} latent spaces.
That is, the dimensionality of the latent space is chosen to be equal to or larger than the
expected intrinsic dimensionality of the data. This choice is essential: if the latent
dimension were fixed to the minimal value from the outset, any apparent dimensional
reduction would be imposed rather than learned.

Let $z = (z_1, \ldots, z_{d_z})$ denote the latent variables of a variational autoencoder
trained on a dataset $\{x^{(i)}\}_{i=1}^N$. For each data point, the encoder defines a
posterior distribution
\begin{equation}
q(z|x^{(i)}) = \prod_{j=1}^{d_z} \mathcal{N}\!\left(z_j \mid \mu_j^{(i)}, \sigma_j^{(i)2}\right),
\end{equation}
where $\mu_j^{(i)}$ and $\sigma_j^{(i)}$ are the mean and standard deviation of the $j$-th
latent variable for event $i$.

If the data contain symmetry-induced redundancies, efficient compression suggests that
only a subset of latent directions should encode significant variation across the dataset.
The remaining directions should be suppressed toward the prior distribution, carrying
little or no information about $x$. This separation between informative and redundant
latent variables is what we refer to as \emph{latent-space self-organization}.

\subsection{Relevance as a Quantitative Diagnostic}

To quantify this effect, one requires a measure that distinguishes variation across events
from uncertainty within a single event. A natural diagnostic exploits the probabilistic
structure of the VAE latent space and compares the spread of latent means across the
dataset to the typical posterior uncertainty.

For each latent variable $z_j$, we define a relevance measure
\begin{equation}
\label{eq:relevance}
\rho_j
\;=\;
\frac{
\mathrm{std}_i\!\left(\mu_j^{(i)}\right)
}{
\left\langle \sigma_j^{(i)} \right\rangle_i
},
\end{equation}
where $\mathrm{std}_i$ denotes the standard deviation over the dataset and
$\langle \cdot \rangle_i$ denotes the average over events.

This ratio has a simple interpretation. The numerator measures how strongly the latent
coordinate $z_j$ varies across different data points, while the denominator measures the
typical uncertainty associated with that coordinate for a single data point. A large value
of $\rho_j$ indicates that the latent variable captures structured, event-to-event
variation that exceeds its intrinsic noise, and is therefore relevant for representing
the dataset. Conversely, values $\rho_j \lesssim \mathcal{O}(1)$ indicate latent variables
whose variation is comparable to or smaller than their uncertainty, suggesting that they
carry little meaningful information.

Importantly, this definition does not assume statistical independence of latent variables
and does not enforce disentanglement. It is therefore well suited to a diagnostic setting,
where the goal is to assess effective dimensionality rather than to recover a unique set
of generative factors.

\subsection{Effective Dimensionality and Hierarchies}

Ordering the latent variables by decreasing relevance,
\begin{equation}
\rho_1 \geq \rho_2 \geq \cdots \geq \rho_{d_z},
\end{equation}
often reveals a pronounced hierarchy. In datasets without strong constraints, several
latent variables may exhibit comparable relevance, reflecting multiple independent
degrees of freedom. In contrast, datasets subject to symmetry constraints frequently show
a sharp drop in $\rho_j$ after a small number of dominant directions.

This hierarchy provides a data-driven estimate of the \emph{effective dimensionality} of
the system. While the precise numerical value of $\rho_j$ depends on architecture and
hyperparameters, the ordering and separation between relevant and irrelevant directions
are typically robust across training runs. As such, the relevance spectrum serves as an
order parameter for symmetry-induced dimensional reduction.

\subsection{Toy Example: Exact Constraint and Dimensional Reduction}

The logic above can be illustrated with a simple example. Consider a dataset of points
embedded in $\mathbb{R}^2$, subject either to no constraint or to an exact constraint
$x_1^2 + x_2^2 = R^2$. In the unconstrained case, the data have two independent degrees of
freedom, and a VAE trained with $d_z \geq 2$ typically exhibits two latent variables with
comparable relevance. In the constrained case, the data lie on a one-dimensional manifold,
and only a single latent direction remains strongly relevant, while the others are
suppressed.

Beyond the relevance spectrum, latent-space organization can be visualized by projecting
the mean latent activations $\langle z_j \rangle$ onto the input space. In symmetry-
constrained datasets, the most relevant latent variable often parametrizes the symmetry
orbit itself, providing a smooth coordinate along the data manifold.

\subsection{Correlations with Physical Observables}

A further layer of interpretation is obtained by studying correlations between relevant
latent variables and physically meaningful combinations of input features. If symmetry
constraints enforce relations such as conservation laws, the dominant latent directions
often correlate with the independent combinations left unconstrained.

For example, in particle collision data, momentum conservation relates the momenta of
final-state particles, reducing the number of independent kinematic variables. When a
VAE is trained on such data, the most relevant latent variables frequently correlate with
differences or invariant combinations of momenta, while directions corresponding to
redundant information are suppressed.

These correlations should not be interpreted as a unique or canonical identification of
physical coordinates. Rather, they provide evidence that the latent representation has
internalized the constraint structure of the data in a nontrivial way.

\subsection{Robustness and Limitations}

The relevance-based diagnostics described above are empirical and subject to limitations.
They depend on successful training, sufficient data coverage of the underlying manifold,
and a clear separation between symmetry-induced redundancy and noise. Approximate or
softly broken symmetries may lead to less pronounced hierarchies, while finite-sample
effects can blur the relevance spectrum.

Nevertheless, when applied judiciously, these diagnostics offer a practical and
interpretable way to assess whether high-dimensional datasets admit a reduced latent
description consistent with underlying symmetries. In the following sections, we apply
this framework to progressively more realistic physical systems, illustrating how exact
and approximate symmetries manifest themselves in latent-space organization.


\section{Case Studies from Physics}
\label{sec:case_studies}

In this section we illustrate how the diagnostic framework introduced above manifests
itself in concrete physical systems, following closely the discussion in Ref.~\cite{Sanz:2025sld}. The goal is not to demonstrate optimal performance
on specific datasets, but to examine how exact and approximate symmetries shape the
organization of latent representations when a variational autoencoder is trained on
physically motivated data. We proceed from simple geometric examples to realistic
scattering processes, increasing the level of complexity and realism at each step.

\subsection{Geometric Constraints and Exact Continuous Symmetries}

\begin{figure}[ht!]
\centering
\includegraphics[width=0.8\textwidth]{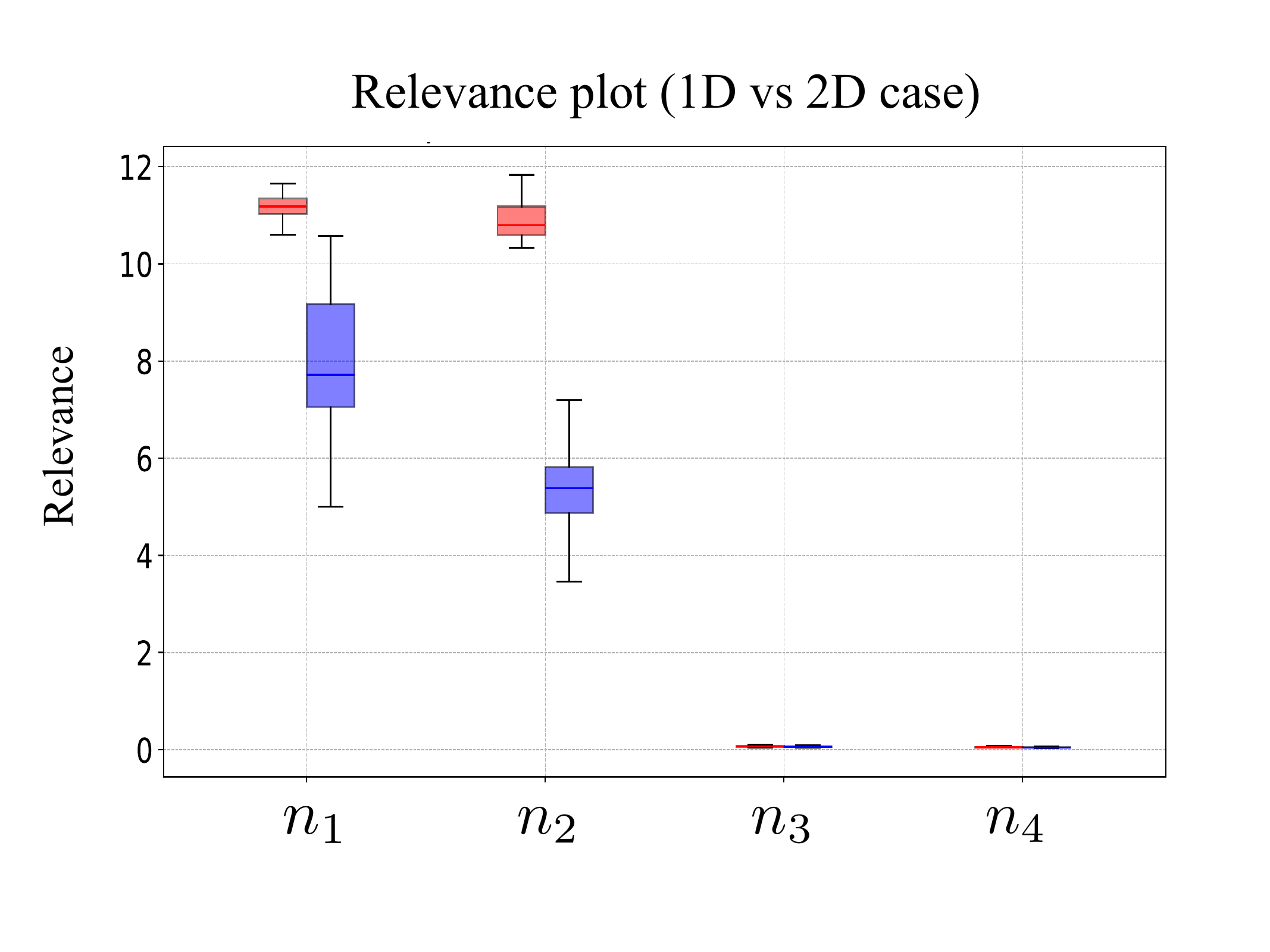}
\includegraphics[width=0.45\textwidth]{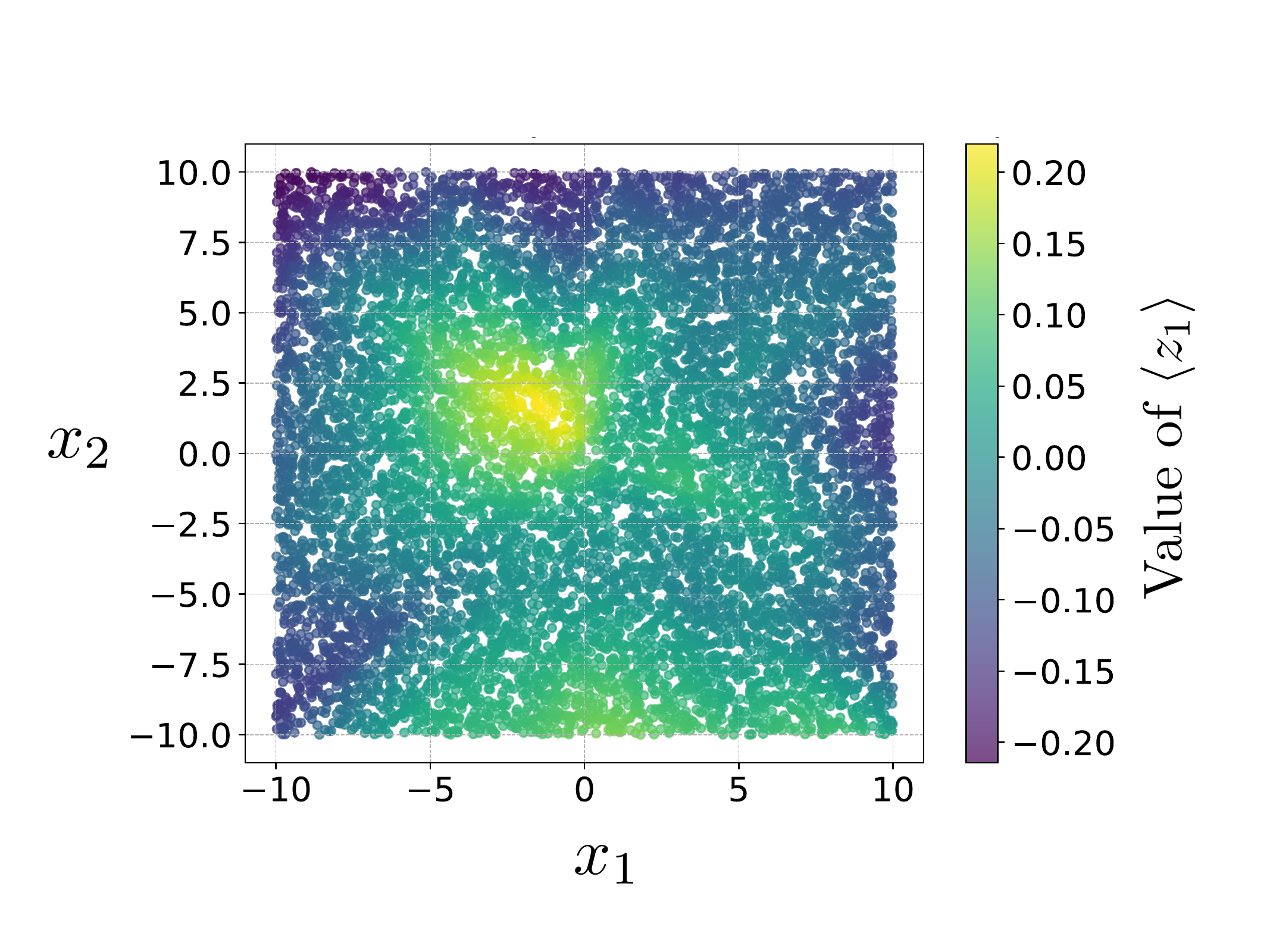}
\includegraphics[width=0.45\textwidth]{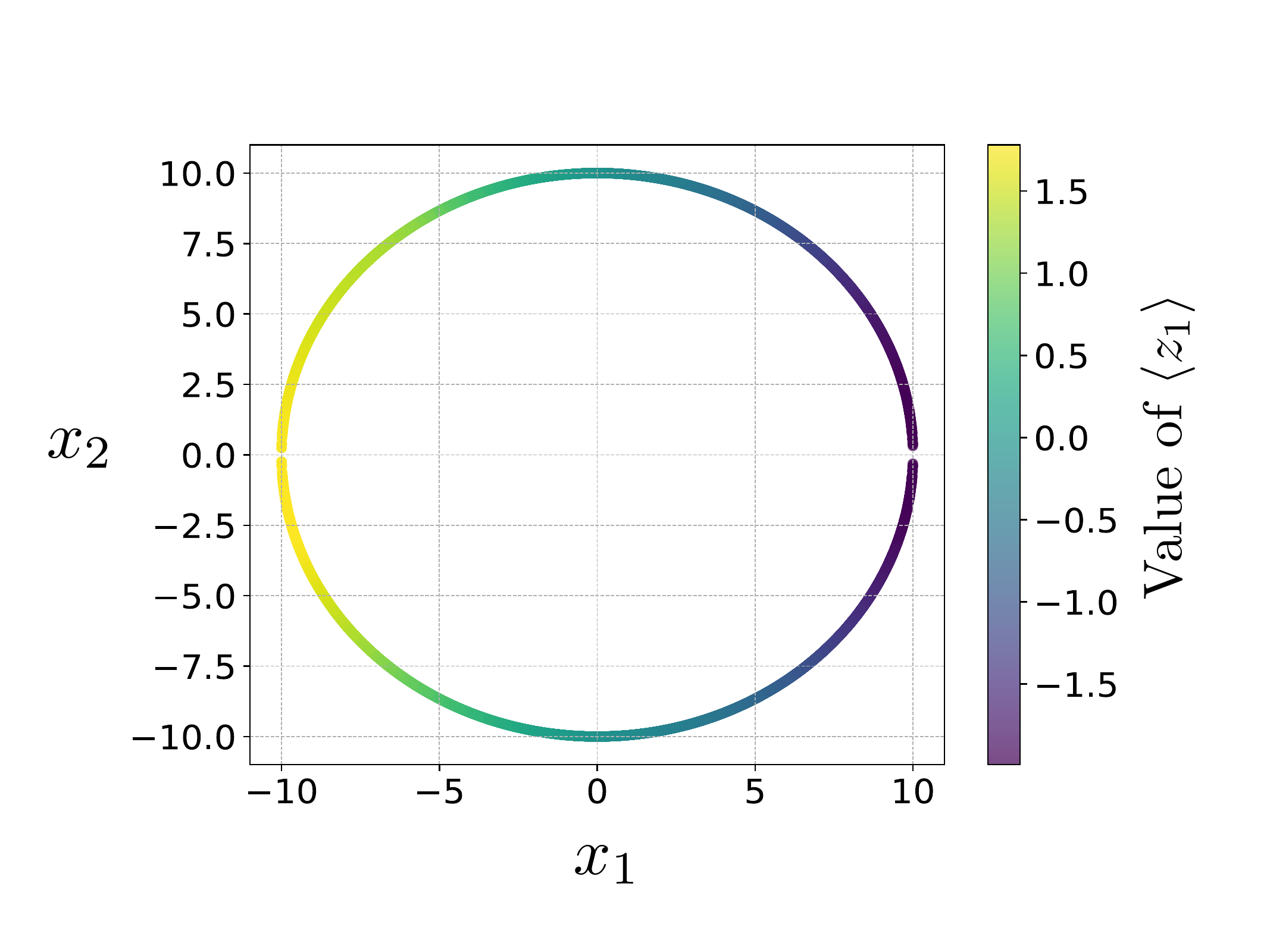}
\caption{
Top plot: Distribution of relevance (as defined in Eq.\ref{eq:relevance}) in the latent variables. In orange, the truly two-dimensional dataset $\mathcal{D}^{\text{2D}}$ and in blue the dataset constrained to a circle $\mathcal{D}^{\text{1D}}$. The latent variables are ordered by decreasing relevance.
Bottom plot: Illustration of latent-space organization for data in 2D (left) and constrained to a circle (right). The dominant latent variable (denoted by $z_1$) provides a smooth coordinate along the symmetry orbit, while remaining latent directions are suppressed. Figures from Ref. \cite{Sanz:2025sld}.
}
\label{fig:circle_latent}
\end{figure}
We begin with simple geometric datasets that provide a controlled environment in which
the role of symmetry is transparent. Consider data points embedded in a two-dimensional
space, generated either without constraints or subject to the exact constraint
$x_1^2 + x_2^2 = R^2$. In the unconstrained case, the data possess two independent degrees
of freedom. In the constrained case, the data lie on a one-dimensional manifold, invariant
under continuous rotations.

When a variational autoencoder is trained on these datasets with a latent dimension
$d_z \geq 2$, a clear distinction emerges.  For the unconstrained dataset, two latent
variables exhibit comparable relevance, reflecting the absence of redundancy. In
contrast, for the constrained dataset, the relevance spectrum shows a pronounced
hierarchy, with a single dominant latent direction and the remaining ones strongly
suppressed. This is shown in the upper panel of Fig.~\ref{fig:circle_latent}. This behavior is stable across independent training runs and choices of
initial conditions.

Beyond the relevance hierarchy, the structure of the dominant latent variable provides
additional insight. Projecting the mean latent activation onto the input space reveals
that this variable parametrizes the angular coordinate along the circle, providing a
smooth ordering of points along the symmetry orbit. Although this coordinate is not
unique—any monotonic reparametrization would serve equally well—it demonstrates that the
latent representation has internalized the effective dimensionality imposed by the
symmetry. This behaviour is shown in the lower panel of Fig.\ref{fig:circle_latent}. All figures are obtained from Ref.\cite{Sanz:2025sld}.

This example serves as a benchmark: when symmetry is exact, global, and cleanly
represented in the data, latent-space self-organization provides a sharp and unambiguous
diagnostic of reduced dimensionality.

\subsection{Lepton Collisions and Momentum Conservation}

We now turn to a physically richer example: electron--positron annihilation into a
muon--antimuon pair, see Fig.\ref{fig:eemumu}. At fixed center-of-mass energy, the final state is fully described by
the three-momenta of the two outgoing leptons. While this yields six observable
components, momentum conservation imposes three exact constraints,
\begin{equation}
\vec{p}_{\mu^+} + \vec{p}_{\mu^-} = 0,
\end{equation}
reducing the number of independent degrees of freedom to three.
\begin{figure}[h!]
\centering
\begin{tikzpicture}
  \begin{feynman}
    \vertex (e-) at (-2,1) {\(e^-(l_1)\)};
    \vertex (e+) at (-2,-1) {\(e^+(l_2)\)};
    \vertex (v1) at (0,0);
    \vertex (v2) at (2,0);
    \vertex (mu-) at (4,1) {\(\mu^-(p_1)\)};
    \vertex (mu+) at (4,-1) {\(\mu^+(p_2)\)};
    
    \diagram* {
      (e-) -- [fermion] (v1) -- [fermion] (e+),
      (v1) -- [photon, momentum=\(\gamma (k)\)] (v2),
      (v2) -- [fermion] (mu-),
      (mu+) -- [fermion] (v2),
    };
  \end{feynman}
\end{tikzpicture}
\caption{Feynman diagram for the process \( e^+ e^- \to \mu^+ \mu^- \) in QED.}
\label{fig:eemumu}
\end{figure}
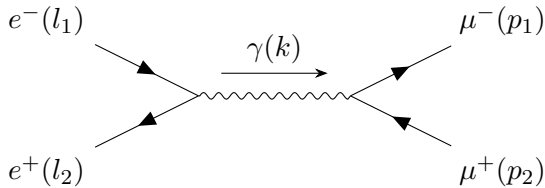
\begin{figure}[ht!]
    \centering
    \includegraphics[width=0.6\textwidth]{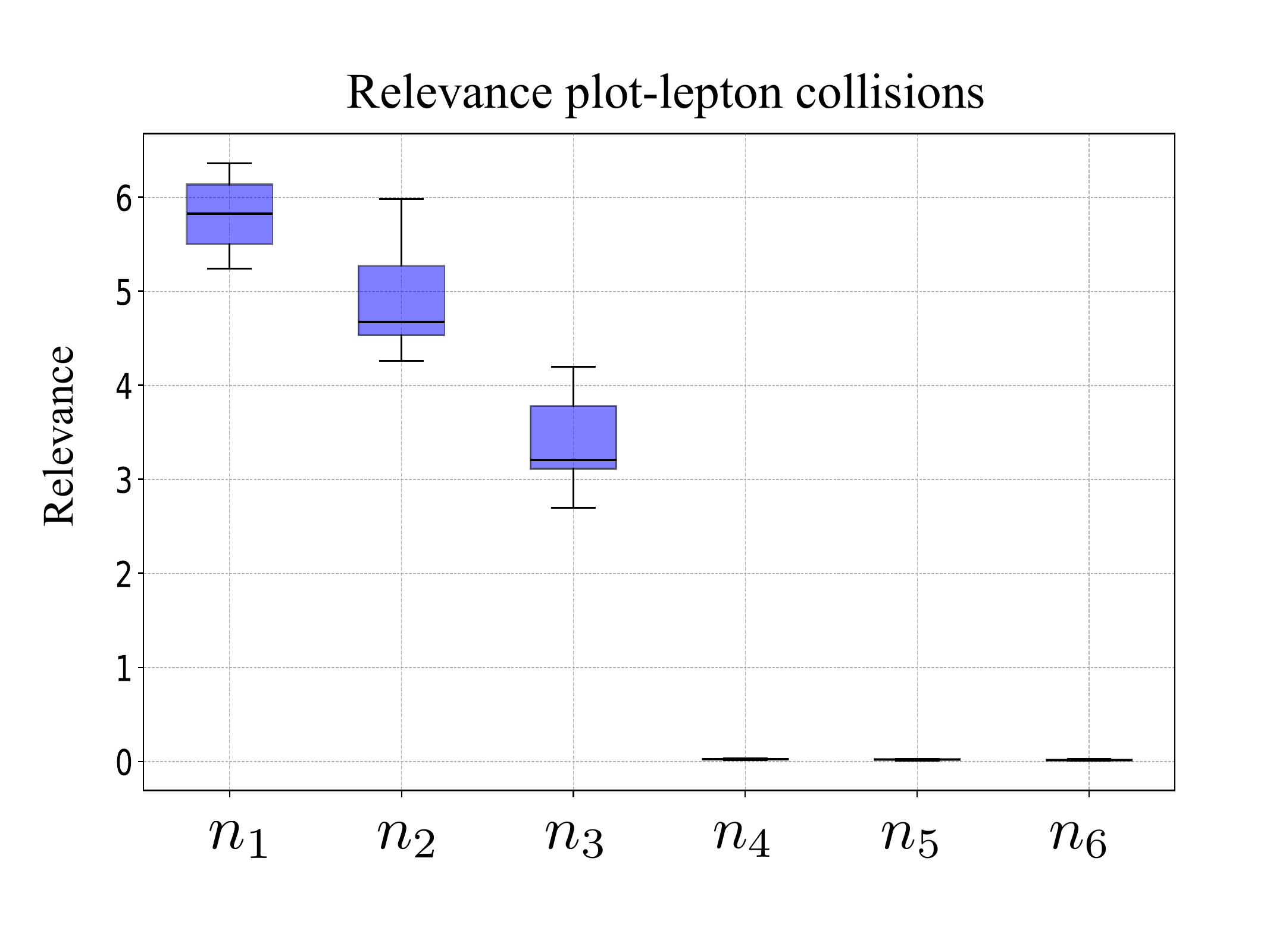}
    \caption{Relevance spectrum and illustrative correlations for lepton collision data.
Three dominant latent directions reflect the effective dimensionality imposed by momentum
conservation. Figure from Ref. \cite{Sanz:2025sld}.}
    \label{fig:relevance_dy}
\end{figure}
\begin{figure}[ht!]
    \centering
    \includegraphics[angle=0,width=0.8\textwidth]{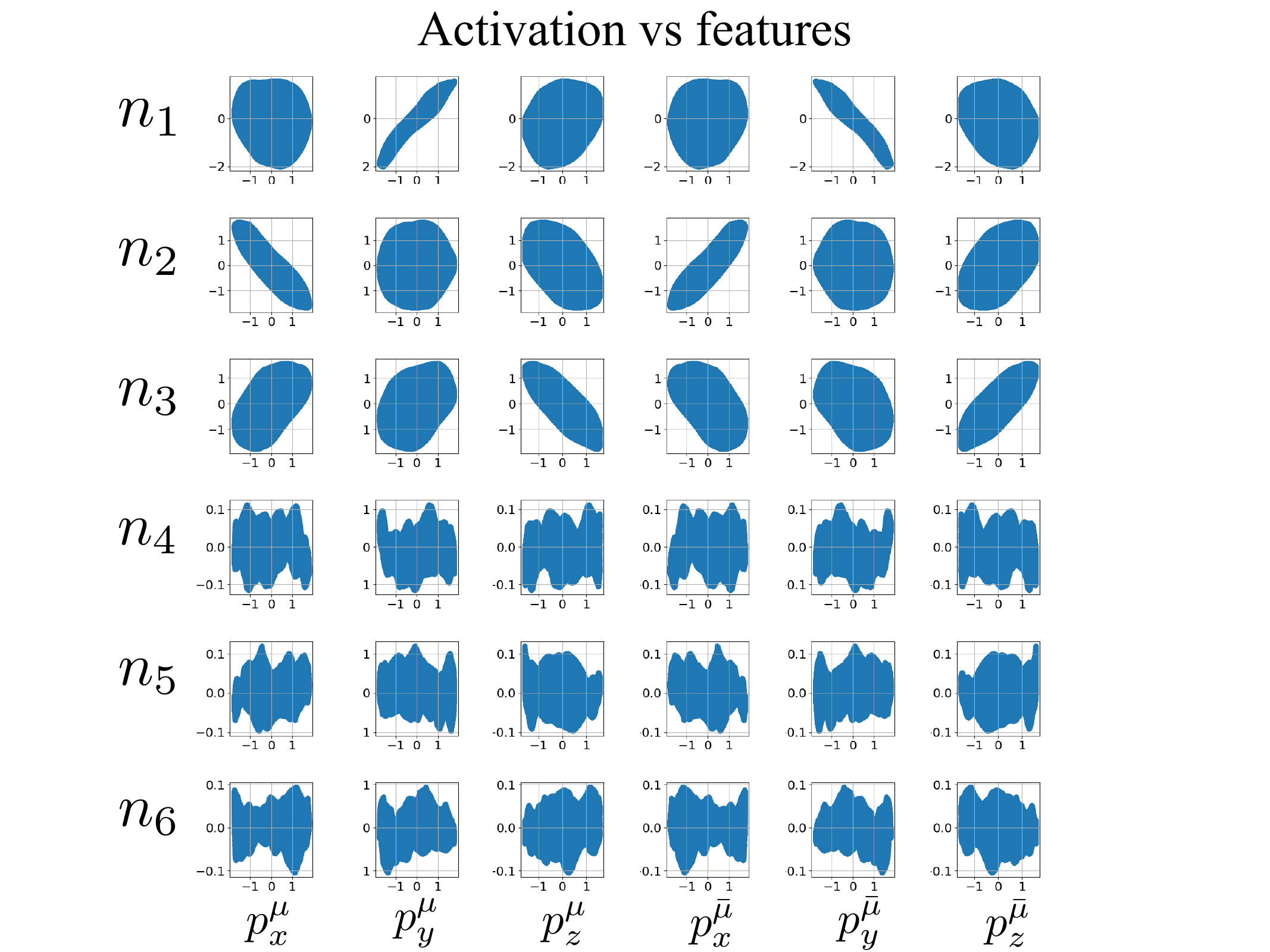}
    \caption{Scatter plots with mean latent activations  as a function of the input kinematic features, for the Drell-Yan dataset. The first three latent variables are strongly correlated with the conserved variable combinations $p_{y}^{\mu} - p_{y}^{\bar{\mu}}$, $p_{x}^{\mu} - p_{x}^{\bar{\mu}}$, and $p_{z}^{\mu} - p_{z}^{\bar{\mu}}$, respectively. Figure from Ref. \cite{Sanz:2025sld}.}
    \label{fig:zmean_dy}
\end{figure}
When a variational autoencoder is trained on such data with a latent space dimension
$d_z \geq 6$, the relevance spectrum again reveals a clear hierarchy. As shown in Fig.\ref{fig:relevance_dy}, only three latent
variables exhibit significant relevance, while the remaining directions collapse toward
the prior. This behavior directly mirrors the dimensional reduction implied by momentum
conservation.

Correlations between the relevant latent variables and the input features provide further
interpretation. As shown in Figure \ref{fig:zmean_dy}, the dominant latent directions typically correlate with independent
combinations of momenta, such as differences between the muon and antimuon momentum
components. Redundant combinations—those fixed by conservation laws—are not encoded in
relevant latent directions.

This case study demonstrates that latent-space self-organization is not limited to toy
models or abstract manifolds. It persists in realistic kinematic datasets, where symmetry
appears as an exact conservation law acting event by event.

\subsection{Hadron Collisions and Approximate Symmetries}

The final example considers Drell--Yan production of muon pairs in proton--proton
collisions. Compared to lepton collisions, the initial state is no longer fully known:
the longitudinal momenta of the incoming partons fluctuate event by event, and the
center-of-mass energy of the hard process is not fixed. As a result, only a subset of the
symmetry constraints present in the electron--positron case survives.

Transverse momentum conservation remains an excellent approximation, while longitudinal
momentum conservation is broken by the unknown parton momentum fractions. Additional
constraints arise from on-shell conditions for the final-state particles and, in selected
kinematic regions, from resonant production through an intermediate vector boson.

\begin{figure}[t]
\centering
\includegraphics[width=0.6\textwidth]{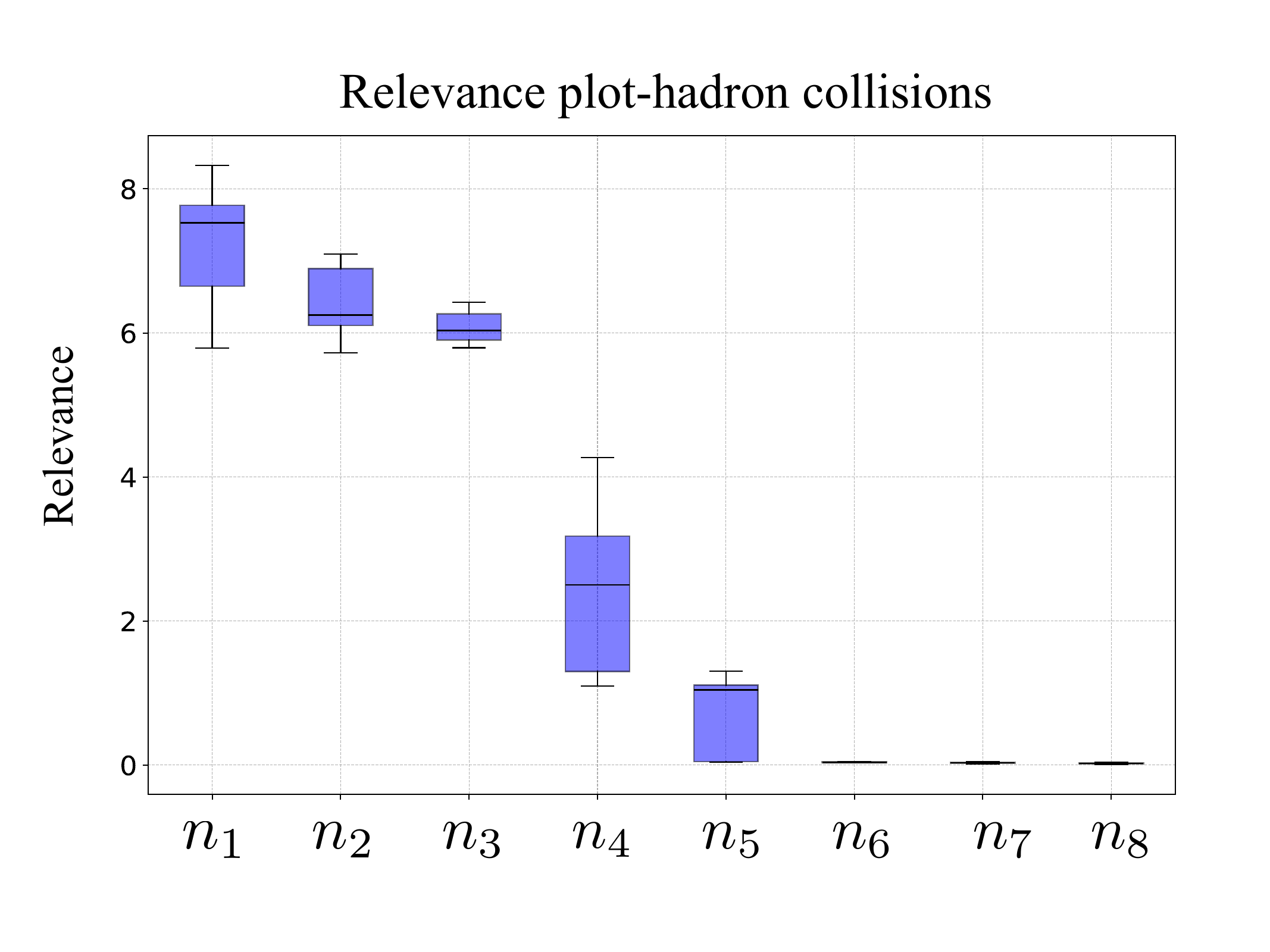}
\caption{Latent relevance spectrum for hadron collision data.
The hierarchy is softer than in cases with exact symmetry, reflecting approximate and
kinematically dependent constraints. Figure from Ref. \cite{Sanz:2025sld}.}
\label{fig:pp_latent}
\end{figure}

When a variational autoencoder is trained on this dataset, the relevance spectrum exhibits
a hierarchy that is less sharp than in the previous cases but remains clearly structured.
As shown in Figure \ref{fig:pp_latent}, a small number of latent variables dominate, reflecting the approximate reduction in
effective dimensionality induced by the surviving constraints. The remaining latent
directions show suppressed but non-negligible relevance, encoding residual variability
associated with broken or softened symmetries. 

This example highlights a crucial point: approximate symmetries do not lead to an abrupt
collapse of latent dimensionality, but rather to graded hierarchies. From a diagnostic
perspective, this behavior is particularly interesting, as it suggests that latent-space
organization can carry information about the degree to which a symmetry is realized in
the data.

\subsection{Lessons from the Case Studies}

Taken together, these examples illustrate a consistent pattern. Exact symmetries lead to
sharp reductions in effective dimensionality and clear latent hierarchies. Approximate or
context-dependent symmetries result in softer but still interpretable structures.
Throughout, the latent representations do not encode symmetry generators explicitly.
Instead, they reflect the informational consequences of symmetry: redundancy, constraint,
and reduced variability.

These observations reinforce the central message of this review. Variational autoencoders,
when used judiciously, can serve as probes of symmetry-induced structure in physical data.
They do not replace analytic reasoning or symmetry principles, but they can provide a
useful diagnostic layer—particularly in complex or high-dimensional settings where the
relevant constraints may not be obvious from first principles.

\section{Theoretical Perspectives on Latent--Symmetry Alignment}
\label{sec:theory_alignment}

The case studies presented in the previous section suggest a recurring empirical pattern:
when data are generated by systems with exact or approximate symmetries, variational
autoencoders trained to balance reconstruction and compression often organize their latent
spaces along symmetry-independent directions. In this section, we examine to what extent
this behavior can be understood from general theoretical considerations, and where its
limitations lie.

\subsection{Symmetry, Redundancy, and Information Compression}

At a conceptual level, the alignment between latent representations and symmetry-induced
structure can be traced back to redundancy. A symmetry identifies multiple configurations
of the data as physically equivalent. From an information-theoretic perspective, such
configurations do not constitute independent messages. Any representation that aims to
encode the data efficiently should therefore avoid allocating independent capacity to
symmetry-related variations.

In a variational autoencoder, this pressure toward compression is made explicit by the
Kullback--Leibler term in the evidence lower bound. Latent directions that do not contribute
significantly to reducing reconstruction error are penalized and driven toward the prior.
If a dataset contains exact constraints—such as conservation laws—then variations along
redundant directions do not improve reconstruction, and these directions are naturally
suppressed. In this sense, symmetry acts indirectly, by shaping the loss landscape rather
than by constraining the model explicitly.

This argument is qualitative but robust. It does not depend on the detailed form of the
symmetry group, nor on the interpretability of the latent variables. It relies only on the
existence of redundancy and on the presence of a compression objective that penalizes
unused degrees of freedom.

\subsection{Linear Limits and the Connection to PCA}

Additional insight can be gained by considering simplified settings in which analytic
results are available. In the linear regime, autoencoders are closely related to principal
component analysis. For a dataset with covariance matrix $\Sigma$, PCA identifies
orthogonal directions that maximize variance. If the data distribution is invariant under
a continuous symmetry, the covariance matrix often reflects this invariance, leading to
degeneracies or reduced rank.

For example, data uniformly distributed on a circle embedded in $\mathbb{R}^2$ have an
isotropic covariance matrix proportional to the identity. Any single projection captures
the same amount of variance, but projecting onto a one-dimensional subspace already
achieves optimal compression. From the PCA perspective, the symmetry manifests itself as a
reduction in effective dimensionality rather than as a preferred direction.

A linear variational autoencoder reproduces this behavior. The reconstruction term
encourages projection onto a low-dimensional subspace, while the KL term penalizes
unnecessary latent dimensions. In this simplified setting, the suppression of redundant
latent variables follows directly from the structure of the data covariance and the form
of the loss. Although nonlinear VAEs operate far beyond this regime, the linear case
provides a useful intuition: symmetry-aligned compression is the generic outcome of
variance-based optimization in the presence of redundancy.

\subsection{Nonlinear Manifolds and Local Coordinate Choices}

In realistic physical datasets, symmetry orbits and constraint surfaces define nonlinear
manifolds embedded in high-dimensional spaces. In such cases, there is no unique global
coordinate system adapted to the symmetry. Any smooth reparametrization of the intrinsic
degrees of freedom is equally valid. Consequently, one should not expect a variational
autoencoder to recover symmetry generators or group parameters in a unique or canonical
form.

Instead, what can be expected is local alignment. In regions of the data manifold where
the geometry is approximately flat, the encoder can learn coordinates that parametrize
variations along the manifold while suppressing transverse directions. Globally, these
coordinates may be distorted or nonlinearly related to physically familiar variables.
From the diagnostic point of view adopted in this review, such distortions are immaterial.
The key signal is the existence of a reduced set of latent directions that carry most of
the variation across the dataset.

This perspective also clarifies why different training runs or architectures may yield
different latent parametrizations while still agreeing on the effective dimensionality.
The alignment is not unique, but the redundancy structure it reflects is.

\subsection{Relation to the Information Bottleneck Principle}

The behavior described above is closely related to the information bottleneck principle,
which frames representation learning as a trade-off between compression of the input and
retention of information relevant for a given task. In the unsupervised setting of a VAE,
the task is reconstruction, and the bottleneck is enforced through the latent prior and
the KL penalty.

From this viewpoint, symmetry-induced redundancy reduces the mutual information between
the data and certain directions of variation. These directions can be discarded with
minimal impact on reconstruction quality. Latent-space self-organization can therefore be
seen as an instance of information bottleneck behavior, where the model preferentially
retains variables that parameterize inequivalent configurations and discards redundant
ones.

It is important to stress, however, that the information bottleneck perspective does not
predict \emph{which} coordinates will be retained, only \emph{how many}. This again
underscores the diagnostic nature of latent-space alignment: it provides information about
dimensionality and redundancy, not about a uniquely defined symmetry representation.

\subsection{Limits of Generality and Absence of Guarantees}

Despite the intuitive arguments above, there is no general theorem guaranteeing that a
variational autoencoder trained on symmetric data will align its latent space with
symmetry directions. Several factors can obstruct this behavior. Finite data coverage,
noise, optimization issues, and architectural choices can all prevent clear latent
hierarchies from emerging. Approximate or softly broken symmetries may lead to weak or
ambiguous signals. In extreme cases, the model may encode redundancy inefficiently, using
multiple latent variables to represent a single effective degree of freedom.

Moreover, the impossibility results for unsupervised disentanglement imply that any claim
of symmetry recovery must be interpreted with care. Without explicit inductive bias,
different latent parametrizations can represent the same data distribution equally well.
The alignment observed in practice is therefore contingent, not fundamental.

For these reasons, latent-space organization should be viewed as evidence \emph{for} the
presence of symmetry-induced structure, not as proof of a specific symmetry or group
action. When combined with physical insight and analytic reasoning, however, it provides
a valuable complementary perspective—particularly in complex systems where the relevant
constraints are not obvious from first principles.

In the final section, we synthesize these insights and discuss how data-driven symmetry
diagnostics may complement traditional approaches in theoretical physics, as well as the
open challenges that remain.


\section{Comparison with Alternative Approaches}
\label{sec:comparison}

The interpretation of latent-space self-organization as a diagnostic of symmetry-induced
structure must be assessed in relation to other methods commonly used for dimensionality
reduction, representation learning, and symmetry handling. In this section we contrast
the relevance-based VAE approach discussed in this review with several widely used
alternatives, emphasizing differences in assumptions, guarantees, and interpretability.

\subsection{Linear Dimensionality Reduction: PCA and Related Methods}

Principal Component Analysis (PCA) remains the standard baseline for identifying reduced
dimensional structure in data~\cite{JolliffePCA,BaldiHornik}. By construction, PCA identifies orthogonal directions of
maximal variance and provides a transparent estimate of effective dimensionality when the
data lie near a linear subspace. In systems with approximate linear constraints, PCA can
indeed reveal rank reduction associated with symmetries or conservation laws.

However, PCA is fundamentally limited to second-order statistics and linear embeddings \cite{RoweisSaulLLE,TenenbaumISOMAP}.
When physical constraints define nonlinear manifolds—such as circular, spherical, or
group-orbit structures—PCA can at best provide a local approximation. While it may signal
reduced rank, it does not yield coordinates adapted to the intrinsic geometry of the data.
Moreover, PCA lacks a probabilistic interpretation that would allow one to distinguish
between genuine structure and noise in a principled way.

In contrast, variational autoencoders generalize this idea to the nonlinear regime and
introduce an explicit notion of latent uncertainty, which is essential for defining
relevance measures that go beyond variance alone.

\subsection{Deterministic Autoencoders}

Deterministic autoencoders extend PCA by allowing nonlinear encoders and decoders.
They can learn compact representations of data lying on nonlinear manifolds and have been
widely used for visualization and compression. From a purely representational standpoint,
they are capable of capturing symmetry-induced dimensional reduction\cite{HintonSalakhutdinov2006,VincentAE}
.

Their main limitation in the present context is interpretability\cite{BengioRepLearning}
. Without a latent prior
or probabilistic structure, the scale, orientation, and usage of latent variables are not
uniquely defined. Redundant directions can persist without penalty, and different training
runs may yield representations that are difficult to compare. As a result, deterministic
autoencoders provide no natural diagnostic for assessing which latent directions are
meaningfully used and which are incidental.

\subsection{Disentanglement-Oriented Variational Models}

A large body of work has focused on encouraging disentangled latent representations\cite{HigginsBetaVAE,KimMnih,ChenTCVAE,Locatello2019}, in
which each latent variable captures an independent factor of variation. Approaches such as
$\beta$-VAE, FactorVAE, and $\beta$-TCVAE modify the standard VAE objective by penalizing
statistical dependence among latent variables.

While these methods can produce more factorized representations in practice, their
limitations are well understood. In the absence of inductive bias or supervision,
disentanglement is not identifiable, and different latent parametrizations can represent
the same data distribution equally well. Moreover, independent factors of variation do not
necessarily correspond to symmetry generators or conserved quantities. A symmetry may
relate variables in a way that is inherently collective rather than separable.

The relevance-based approach adopted here does not aim to produce disentanglement.
Instead, it asks a weaker but more robust question: how many latent directions are
effectively used, and does this number reflect symmetry-induced redundancy in the data?

\subsection{Equivariant and Group-Structured Models}

When the relevant symmetry group is known, equivariant neural networks and group-structured
generative models offer strong guarantees \cite{CohenWelling2016,BronsteinGDL}
. By construction, their latent variables
transform in prescribed ways under group actions, ensuring alignment with symmetry
directions. Such models are particularly powerful in domains where symmetry principles are
firmly established, such as Lorentz invariance in high-energy physics \cite{BogatskiyPELICAN,HaoLorentzAE,KondorTrivedi}
or permutation
invariance in set-based data.

The limitation of this approach is its reliance on prior knowledge. The symmetry group,
its representation, and its action on the data must be specified in advance. This makes
equivariant models ill-suited to exploratory settings where symmetries are unknown,
approximate, or emergent. In these cases, imposing an incorrect symmetry can bias the
analysis and obscure physically relevant effects.

The diagnostic approach discussed in this review is complementary. It does not compete
with equivariant modeling, but rather provides a way to assess whether symmetry-aware
architectures are warranted in the first place.

\subsection{Normalizing Flows and Diffusion Models}

Modern generative models based on normalizing flows or diffusion processes achieve
state-of-the-art performance in density estimation and sample generation \cite{RezendeFlows,DinhRealNVP,KingmaGlow,HoDiffusion,SongScore}. Their latent
spaces, however, are typically optimized for likelihood rather than interpretability \cite{BrehmerINN}.
While these models can faithfully reproduce complex data distributions, the internal
representation of symmetry-related structure is often implicit and difficult to extract.

In principle, symmetry diagnostics could be applied post hoc to these models, but the
absence of a simple probabilistic notion of latent relevance complicates the analysis.
For the purpose of probing effective dimensionality and redundancy, VAEs occupy a useful
middle ground between expressive power and interpretability.

\subsection{Summary of Comparative Features}

To summarize the discussion, Table~\ref{tab:comparison} contrasts the main approaches
along dimensions relevant for symmetry diagnostics~\cite{BronsteinGDL,Locatello2019,KingmaWelling}.

\begin{table}[t]
\centering
\begin{tabular}{lcccc}
\hline
Method & Nonlinear & Probabilistic & Symmetry Required & Diagnostic Use \\
\hline
PCA & No & No & No & Limited \\
Deterministic AE & Yes & No & No & Limited \\
Disentangled VAE & Yes & Yes & Implicit & Partial \\
Equivariant Models & Yes & Optional & Yes & Enforced \\
VAE (this work) & Yes & Yes & No & Diagnostic \\
\hline
\end{tabular}
\caption{Comparison of common representation-learning approaches from the perspective of
symmetry diagnostics.}
\label{tab:comparison}
\end{table}

Overall, no single method provides a universal solution. Equivariant models are optimal
when symmetries are known; disentanglement-oriented models are useful when independent
factors are expected; and linear methods remain valuable baselines. The relevance-based
VAE framework reviewed here occupies a distinct niche: it offers a principled and
interpretable diagnostic for detecting symmetry-induced structure in data, without
requiring prior knowledge of the symmetry itself.


\section{Outlook and Open Problems}
\label{sec:outlook}

The case studies and theoretical considerations reviewed in this article suggest that
latent-space self-organization in generative models can provide useful diagnostics of
symmetry-induced structure in physical data. At the same time, they highlight clear
limitations and open questions that must be addressed before such methods can be
systematically integrated into the theoretical physics toolkit. In this section, we
outline several directions where further progress is both necessary and promising.

\subsection{Approximate, Broken, and Emergent Symmetries}

Most symmetries encountered in realistic physical systems are not exact. They may be
explicitly broken by interactions, softly violated by external conditions, or emerge only
in restricted regimes. Understanding how such situations are reflected in latent-space
organization remains an open challenge. While exact symmetries tend to produce sharp
hierarchies in latent relevance, approximate symmetries lead to graded spectra whose
interpretation is less straightforward.

A systematic study of how symmetry-breaking scales, noise levels, and finite-sample
effects map onto latent relevance hierarchies would be valuable. Such studies could help
distinguish genuine physical symmetry breaking from artifacts of limited data or model
capacity, and may provide a quantitative way to assess the degree to which a symmetry is
realized in experimental data.

\subsection{Unknown Symmetries and Exploratory Diagnostics}

One of the main motivations for a diagnostic, rather than prescriptive, approach to
symmetry is the possibility of encountering unknown or unexpected structure. In complex
systems—ranging from many-body dynamics to astrophysical or cosmological observations—it
is often unclear which symmetries, if any, should be imposed at the outset.

In such settings, latent-space diagnostics may serve as a preliminary exploratory tool.
Evidence for reduced effective dimensionality or structured redundancy could motivate the
introduction of symmetry-aware models, effective theories, or analytic ansätze at a later
stage. Conversely, the absence of such signals may indicate that symmetry-based modeling
is not appropriate, or that relevant constraints lie outside the observed feature space.

\subsection{Connections to Effective Field Theory and Model Building}

The logic underlying effective field theory is deeply connected to symmetry and
dimensional reduction: irrelevant degrees of freedom are integrated out, while the
remaining dynamics are organized by symmetry principles. From this perspective, data-
driven diagnostics of effective dimensionality may provide a complementary angle on EFT
construction, particularly in regimes where the appropriate degrees of freedom are not
obvious.

While current methods do not identify operators or symmetry generators explicitly, they
may help delineate the boundary between relevant and redundant structures in data. Future
work combining latent-space diagnostics with symbolic regression, sparse modeling, or
operator inference may help bridge the gap between purely data-driven representations and
analytic theoretical frameworks.

\subsection{Architectural Biases and Hybrid Approaches}

Another open direction concerns the controlled introduction of inductive bias. While this
review has focused on standard variational autoencoders precisely because they do not
enforce symmetry by construction, hybrid approaches are worth exploring. One could imagine
models that remain agnostic about the precise symmetry group, but are biased toward
low-dimensional, structured latent representations through weak regularization or
hierarchical priors.

Such approaches may strike a balance between flexibility and interpretability, allowing
symmetry-related structure to emerge when supported by the data, while avoiding the risks
associated with enforcing incorrect symmetries.

\subsection{From Diagnostics to Practice}

Finally, translating latent-space diagnostics into practical tools for data analysis
requires careful validation. Sensitivity to architecture, hyperparameters, and training
procedures must be quantified, and diagnostic criteria must be calibrated against known
benchmarks. Without such validation, there is a risk of over-interpreting latent structure
as physically meaningful when it is not.

Addressing these challenges will require close interaction between machine learning,
theoretical modeling, and domain expertise. The potential payoff, however, is significant:
a set of data-driven probes that can reveal hidden structure before committing to specific
theoretical assumptions.


\section{Conclusions}
\label{sec:conclusions}

Symmetry has long served as a guiding principle in theoretical physics, shaping our
understanding of fundamental interactions and constraining the space of viable models.
At the same time, modern machine learning has introduced powerful new tools for extracting
structure from complex, high-dimensional data. This review has examined the intersection
of these two perspectives, focusing on the extent to which symmetry-induced constraints
and redundancies can be diagnosed directly from data through representation learning.

Rather than emphasizing architectures that enforce symmetry by construction, we have
adopted a diagnostic viewpoint. Using variational autoencoders as a concrete example, we
have shown how the presence of exact or approximate symmetries can manifest itself through
self-organization of latent spaces, leading to hierarchies of relevance and effective
dimensional reduction. Case studies ranging from simple geometric systems to realistic
particle physics processes illustrate how conservation laws and kinematic constraints are
reflected in learned representations.

We have also stressed the limitations of this approach. Latent-space alignment with
symmetry directions is neither guaranteed nor unique, and cannot be interpreted as the
recovery of symmetry generators in a strict sense. Impossibility results for unsupervised
disentanglement, finite data effects, and model dependence all impose fundamental
constraints on what can be inferred. For these reasons, data-driven symmetry diagnostics
should be viewed as complementary to, rather than a replacement for, traditional analytic
reasoning.

Within these limits, however, the diagnostic framework reviewed here offers a useful new
layer of analysis. By probing effective dimensionality and redundancy directly from data,
it can help identify when symmetry-based descriptions are appropriate, guide the choice of
modeling assumptions, and highlight regimes where unexpected structure may be present.
As data volumes and complexity continue to grow across many areas of physics, such tools
may become increasingly valuable in navigating the interface between empirical evidence
and theoretical interpretation.

In this sense, artificial intelligence does not challenge the central role of symmetry in
physics. Instead, it provides new ways of interrogating how symmetry leaves its imprint on
data—and of discovering when that imprint is present, subtle, or absent altogether.

\section*{Acknowledgments}

This work
is supported by the grants PID2023-148162NB-C21 and
 CEX2023-001292-S from the Ministerio de Ciencia, Innovacion y Universidades.
\bibliographystyle{jhep}
\bibliography{ai_symmetries}

@article{Noether1918,
  author = {Noether, Emmy},
  title = {Invariante Variationsprobleme},
  journal = {Nachrichten von der Gesellschaft der Wissenschaften zu G\"ottingen},
  year = {1918},
  pages = {235--257}
}

@book{WeinbergQFT1,
  author = {Weinberg, Steven},
  title = {The Quantum Theory of Fields, Volume I},
  publisher = {Cambridge University Press},
  year = {1995}
}

@book{WeinbergQFT2,
  author = {Weinberg, Steven},
  title = {The Quantum Theory of Fields, Volume II},
  publisher = {Cambridge University Press},
  year = {1996}
}

@article{BengioRepLearning,
  author = {Bengio, Yoshua and Courville, Aaron and Vincent, Pascal},
  title = {Representation Learning: A Review and New Perspectives},
  journal = {IEEE Transactions on Pattern Analysis and Machine Intelligence},
  volume = {35},
  year = {2013},
  pages = {1798--1828}
}

@book{GoodfellowDL,
  author = {Goodfellow, Ian and Bengio, Yoshua and Courville, Aaron},
  title = {Deep Learning},
  publisher = {MIT Press},
  year = {2016}
}

@article{LeCunCNN,
  author = {LeCun, Yann and Bottou, L\'eon and Bengio, Yoshua and Haffner, Patrick},
  title = {Gradient-Based Learning Applied to Document Recognition},
  journal = {Proceedings of the IEEE},
  volume = {86},
  year = {1998},
  pages = {2278--2324}
}

@inproceedings{CohenWelling2016,
  author = {Cohen, Taco and Welling, Max},
  title = {Group Equivariant Convolutional Networks},
  booktitle = {Proceedings of the 33rd International Conference on Machine Learning},
  year = {2016}
}

@article{BronsteinGDL,
  author = {Bronstein, Michael M. and Bruna, Joan and Cohen, Taco and Veli{\v c}kovi{\'c}, Petar},
  title = {Geometric Deep Learning: Grids, Groups, Graphs, Geodesics, and Gauges},
  journal = {arXiv:2104.13478},
  year = {2021}
}

@article{HigginsBetaVAE,
  author = {Higgins, Irina and et al.},
  title = {$\beta$-VAE: Learning Basic Visual Concepts with a Constrained Variational Framework},
  journal = {ICLR},
  year = {2017}
}

@article{InfoGAN,
  author = {Chen, Xi and et al.},
  title = {InfoGAN: Interpretable Representation Learning by Information Maximizing Generative Adversarial Nets},
  journal = {NeurIPS},
  year = {2016}
}

@article{Locatello2019,
  author = {Locatello, Francesco and et al.},
  title = {Challenging Common Assumptions in the Unsupervised Learning of Disentangled Representations},
  journal = {ICML},
  year = {2019}
}

@article{KingmaWelling,
  author = {Kingma, Diederik P. and Welling, Max},
  title = {Auto-Encoding Variational Bayes},
  journal = {ICLR},
  year = {2014}
}

@article{BogatskiyPELICAN,
  author = {Bogatskiy, Alexander and Hoffman, Thomas and Miller, David W. and Offermann, Julian T.},
  title = {PELICAN: Permutation Equivariant and Lorentz Invariant or Covariant Aggregator Network for Particle Physics},
  journal = {arXiv:2211.00454},
  year = {2022}
}

@article{BaldiHornik,
  author = {Baldi, Pierre and Hornik, Kurt},
  title = {Neural Networks and Principal Component Analysis: Learning from Examples without Local Minima},
  journal = {Neural Networks},
  volume = {2},
  year = {1989},
  pages = {53--58}
}

@article{KimMnih,
  author = {Kim, Hyunjik and Mnih, Andriy},
  title = {Disentangling by Factorising},
  journal = {ICML},
  year = {2018}
}

@book{JolliffePCA,
  author = {Jolliffe, Ian T.},
  title = {Principal Component Analysis},
  publisher = {Springer},
  edition = {2},
  year = {2002}
}

@article{RoweisSaulLLE,
  author = {Roweis, Sam T. and Saul, Lawrence K.},
  title = {Nonlinear Dimensionality Reduction by Locally Linear Embedding},
  journal = {Science},
  volume = {290},
  year = {2000},
  pages = {2323--2326}
}

@article{TenenbaumISOMAP,
  author = {Tenenbaum, Joshua B. and de Silva, Vin and Langford, John C.},
  title = {A Global Geometric Framework for Nonlinear Dimensionality Reduction},
  journal = {Science},
  volume = {290},
  year = {2000},
  pages = {2319--2323}
}

@article{HintonSalakhutdinov2006,
  author = {Hinton, Geoffrey E. and Salakhutdinov, Ruslan},
  title = {Reducing the Dimensionality of Data with Neural Networks},
  journal = {Science},
  volume = {313},
  year = {2006},
  pages = {504--507}
}

@article{VincentAE,
  author = {Vincent, Pascal and et al.},
  title = {Extracting and Composing Robust Features with Denoising Autoencoders},
  journal = {ICML},
  year = {2008}
}

@article{RezendeFlows,
  author = {Rezende, Danilo J. and Mohamed, Shakir},
  title = {Variational Inference with Normalizing Flows},
  journal = {ICML},
  year = {2015}
}

@article{DinhRealNVP,
  author = {Dinh, Laurent and Sohl-Dickstein, Jascha and Bengio, Samy},
  title = {Density Estimation using Real NVP},
  journal = {ICLR},
  year = {2017}
}

@article{KingmaGlow,
  author = {Kingma, Diederik P. and Dhariwal, Prafulla},
  title = {Glow: Generative Flow with Invertible 1x1 Convolutions},
  journal = {NeurIPS},
  year = {2018}
}

@article{HoDiffusion,
  author = {Ho, Jonathan and Jain, Ajay and Abbeel, Pieter},
  title = {Denoising Diffusion Probabilistic Models},
  journal = {NeurIPS},
  year = {2020}
}

@article{Barenboim:2021vzh,
    author = "Barenboim, Gabriela and Hirn, Johannes and Sanz, Veronica",
    title = "{Symmetry meets AI}",
    eprint = "2103.06115",
    archivePrefix = "arXiv",
    primaryClass = "cs.LG",
    doi = "10.21468/SciPostPhys.11.1.014",
    journal = "SciPost Phys.",
    volume = "11",
    pages = "014",
    year = "2021"
}

@article{barenboim2024exploring,
  title={Exploring how a generative AI interprets music},
  author={Barenboim, Gabriela and Debbio, Luigi Del and Hirn, Johannes and Sanz, Ver{\'o}nica},
  journal={Neural Computing and Applications},
  volume={36},
  number={27},
  pages={17007--17022},
  year={2024},
  publisher={Springer}
}

@article{SongScore,
  author = {Song, Yang and et al.},
  title = {Score-Based Generative Modeling through Stochastic Differential Equations},
  journal = {ICLR},
  year = {2021}
}

@article{BrehmerINN,
  author = {Brehmer, Johann and Louppe, Gilles and Pavez, Juan and Cranmer, Kyle},
  title = {Mining Gold from Implicit Models to Improve Likelihood-Free Inference},
  journal = {PNAS},
  volume = {117},
  year = {2020},
  pages = {5242--5249}
}

@article{KondorTrivedi,
  author = {Kondor, Risi and Trivedi, Shubhendu},
  title = {On the Generalization of Equivariance and Convolution in Neural Networks to the Action of Compact Groups},
  journal = {ICML},
  year = {2018}
}

@article{ChenTCVAE,
  author = {Chen, Ricky T.Q. and et al.},
  title = {Isolating Sources of Disentanglement in Variational Autoencoders},
  journal = {NeurIPS},
  year = {2018}
}

@article{Sanz:2025sld,
    author = "Sanz, Veronica",
    title = "{Learning symmetries in datasets}",
    eprint = "2504.05174",
    archivePrefix = "arXiv",
    primaryClass = "cs.LG",
    month = "4",
    year = "2025"
}

@article{HaoLorentzAE,
  author = {Hao, Zhiqiang and Kansal, Raghav and Duarte, Javier and Chernyavskaya, Nelya},
  title = {Lorentz Group Equivariant Autoencoders},
  journal = {Eur. Phys. J. C},
  volume = {83},
  year = {2023},
  pages = {485}
}

@article{JetCLR,
  author = {Dillon, Barry M. and Kasieczka, Gregor and Olischl{\"a}ger, Helge and Plehn, Tilman and Sorrenson, Peter and Vogel, Lukas},
  title = {Symmetries, Safety, and Self-Supervision},
  journal = {SciPost Physics},
  volume = {12},
  year = {2022},
  pages = {188}
}

@article{DillonSymmetrySafety,
  author = {Dillon, Barry M. and Plehn, Tilman and Sauer, Christopher and Sorrenson, Peter},
  title = {Better Latent Spaces for Better Autoencoders},
  journal = {SciPost Physics},
  volume = {11},
  year = {2021},
  pages = {061}
}

@article{ChenSimCLR,
  author = {Chen, Ting and et al.},
  title = {A Simple Framework for Contrastive Learning of Visual Representations},
  journal = {ICML},
  year = {2020}
}

@article{GrillBYOL,
  author = {Grill, Jean-Bastien and et al.},
  title = {Bootstrap Your Own Latent: A New Approach to Self-Supervised Learning},
  journal = {NeurIPS},
  year = {2020}
}

@book{PeskinSchroeder,
  author = {Peskin, Michael E. and Schroeder, Daniel V.},
  title = {An Introduction to Quantum Field Theory},
  publisher = {Westview Press},
  year = {1995}
}

@book{GeorgiEFT,
  author = {Georgi, Howard},
  title = {Weak Interactions and Modern Particle Theory},
  publisher = {Benjamin/Cummings},
  year = {1984}
}

\end{document}